\documentstyle[floats,aps,epsf,prb,preprint]{revtex}
\begin{document}
\draft
\tighten

%.......................................................................

\title{Theory of PbTiO$_3$, BaTiO$_3$, and SrTiO$_3$ Surfaces}
\author{B.~Meyer, J.~Padilla, and David Vanderbilt}
\address{Department of Physics and Astronomy, Rutgers University,
Piscataway, New Jersey 08855-0849}

\date{March 1, 1999}
\maketitle

\begin{abstract}
First-principles total-energy calculations are carried out for
(001) surfaces of the cubic perovskite ATiO$_3$ compounds
PbTiO$_3$, BaTiO$_3$, and SrTiO$_3$.  Both AO-terminated and
TiO$_2$-terminated surfaces are considered, and fully-relaxed
atomic configurations are determined.  In general, BaTiO$_3$ and
SrTiO$_3$ are found to have a rather similar behavior, while
PbTiO$_3$ is different in many respects because of the partially
covalent character of the Pb--O bonds.  PbTiO$_3$ and BaTiO$_3$
are ferroelectrics, and the influence of the surface upon the
ferroelectric distortions is studied for the case of a tetragonal
ferroelectric distortion parallel to the surface.  The surface
relaxation energies are found to be substantial, i.e., many times
larger than the bulk ferroelectric well depth.  Nevertheless, the
influence of the surface upon the ferroelectric order parameter
is modest, and is qualitatively as well as quantitatively
different for the two materials.  Surface energies and electronic
properties are also computed.  It is found that for BaTiO$_3$ and
SrTiO$_3$ surfaces, both AO-terminated and TiO$_2$-terminated
surfaces can be thermodynamically stable, whereas for PbTiO$_3$
only the PbO surface termination is stable.
\end{abstract}

\narrowtext

% -------------------------------------------------------------------

\section{Introduction}
\label{sec:introduction}

The surfaces of insulating cubic perovskite materials such as
PbTiO$_3$, BaTiO$_3$, and SrTiO$_3$ are of interest from several
points of view.
First, some of these materials (notably SrTiO$_3$) are very
widely used as substrates for growth of other oxide materials
(e.g., layered high-$T_c$ superconductors and ``colossal
magnetoresistance'' materials).
Second, this class of materials is of enormous importance for
actual and potential applications that make use of their unusual
piezoelectric, ferroelectric, and dielectric properties (e.g.,
for piezoelectric transducers, non-volatile memories, and
wireless communications applications, respectively).  Many of
these applications are increasingly oriented towards thin-film
geometries, where surface properties are of growing importance.
Third, the bulk materials display a variety of
structural phase transitions; the ferroelectric (FE) structural
phases are of special interest, but antiferroelectric (AFE) or
antiferrodistortive (AFD) transitions can also take
place.\cite{lines}  It is then of considerable fundamental
interest to consider how these structural distortions couple to
the surface, e.g., whether the presence of the surface acts to
enhance or suppress the structural distortion.  The
ferroelectric properties are well known to degrade in
thin-film\cite{tsai} and particulate\cite{niepce} geometries,
and it is very important to understand whether such behavior is
intrinsic to the presence of a surface, or whether it arises from
extrinsic factors such as compositional non-uniformities or
structural defects in the surface region.
Finally, the cubic perovskites can serve as model systems
for the study of transition-metal oxide surfaces more
generally.\cite{cox}

In the last decade, there has been a surge of activity in the
application of first-principles computational methods based on
density-functional theory (DFT) to the study of the bulk
properties, and especially the ferroelectric transitions, in bulk
perovskite oxides.  (For a recent review, see
Ref.~\onlinecite{dv-co} or \onlinecite{dv-korea}.)  The importance
of these methods was recently underlined by the award of the Nobel
Prize in Chemistry to Walter Kohn, the primary originator of DFT.
In the materials theory community, these methods have been widely
used for two decades to predict properties of
semiconductors and simple metals.  However, recent advances
in computational algorithms and computer power now allow these
methods to be applied to more complex materials (e.g., perovskites)
and more complex geometries (e.g., defects and surfaces).
In particular, pioneering studies of
BaTiO$_3$ \cite{cohen1,cohen2,pad-ba}
and SrTiO$_3$ \cite{kimura,pad-sr,li-et-al}
surfaces have recently appeared.

Experimental investigations of the surface structure of cubic
perovskites have not been very extensive.  Such studies are
hindered by the difficulties of preparing clean and defect-free
surfaces, and of overcoming charging effects associated with many
experimental probes.  Even for SrTiO$_3$, the best-studied of
these surfaces, there is a disappointing level of agreement among
experimental results \cite{hikita,bickel,naito,kita} and between
experiment and theory.\cite{pad-sr}  We are not aware of
comparable studies of BaTiO$_3$ and PbTiO$_3$ surfaces.

The purpose of the present contribution is to present new theoretical
work on the structural properties of the PbTiO$_3$ (001) surface, and
to compare and contrast these results with the previous work
of our group on BaTiO$_3$ and SrTiO$_3$ surfaces.\cite{pad-ba,pad-sr}
As regards bulk properties, lead-based compounds such as PbTiO$_3$
and PbZrO$_3$ are known to behave quite differently from
alkaline-earth based perovskites such as BaTiO$_3$ and SrTiO$_3$.
Previous theoretical work has shown that the FE distortion is typically
larger and that Pb atoms participate much more strongly in (and
sometimes even dominate) the FE distortion, compared with non-Pb
perovskites.\cite{coh-nature,coh-krak-2,king,gar-pr,rabe-pbzr}
Moreover, the Pb-based compounds are generally more susceptible
to more complex AFD and AFE instabilities involving tilting of
the oxygen octahedra,\cite{gar-pr,rabe-pbzr,z-sr,singh-pbzr}
and the ground-state structures often involve the formation of some
quite short Pb--O bonds.\cite{singh-pbzr,lb-pzt,lb-hetero,egami}
All of these effects point to a strong and active involvement
of the Pb atoms in the bonding, most naturally interpreted in terms
of the formation of partially covalent Pb--O bonds with the closest
oxygen neighbors.
Finally, a focus on Pb-based materials is motivated by the fact that
these are the leading candidates for many practical piezoelectric
and switching applications, especially in the form of solid solutions
such as
PZT (PbZr$_x$Ti$_{1-x}$O$_3$),
PMN (PbMg$_{1/3}$Nb$_{2/3}$O$_3$), and
PZN (PbZn$_{1/3}$Nb$_{2/3}$O$_3$).

The manuscript is organized as follows.  Section II contains a
brief account of the technical details of the work, including
the theoretical methods used, the slab geometries studied,
and the formulation of the surface energy.  In Sec.~III we
present the computed structural relaxations of the PbTiO$_3$
surfaces, and compare these to the previous results on
BaTiO$_3$ and SrTiO$_3$ surfaces. Additionally, we discuss the
surface energetics (surface energies and surface relaxation
energies), and point out some characteristic differences in the
surface electronic structure of the three compounds. Finally,
the paper ends with a summary in Sec.~\ref{sec:summary}.

% ----------------------------------------------------------------
\section{Preliminaries}
% ----------------------------------------------------------------

% ------------------
\subsection{Theoretical Methods}
% ------------------

We carried out self-consistent plane-wave pseudopotential
calculations within Kohn-Sham density-functional theory
using a conjugate-gradient technique.\cite{king}
Exchange and correlation were treated using the Ceperley-Alder
form.\cite{ceper}  Vanderbilt ultrasoft pseudopotentials were
employed,\cite{vand-usp} with semicore Pb $5d$, Ba $5s$ and $5p$,
Sr $4s$ and $4p$, and Ti $3s$ and $3p$ orbitals included as
valence states.  A plane-wave cutoff of 25 Ry has been used
throughout.  Relaxations of atomic coordinates are iterated
until the forces are less than 0.01~eV/\AA.
Justification of the convergence and accuracy
of this approach can be found in the previously published
work.\cite{pad-ba,pad-sr,king}

% ------------------
\subsection{Surface and Slab Geometries}
\label{sec:geometry}
% ------------------

In this work we consider only II-IV cubic perovskites, i.e.,
ABO$_3$ perovskites in which atoms A and B are divalent and
tetravalent, respectively.  In this case, two non-polar (001)
surface terminations are possible: the AO--terminated surface,
and the BO$_2$--terminated surface.

\begin{figure}
$\phantom{dummy}$
\vfill
\epsfysize=3.0in
\centerline{\epsffile{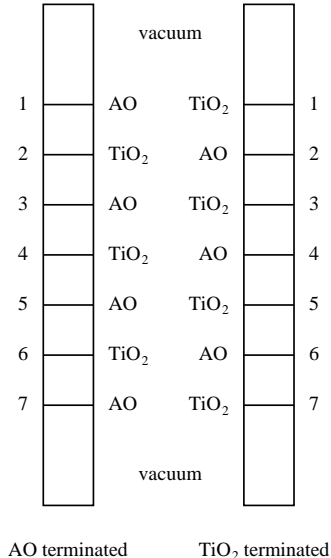}}
\vfill
\caption{\label{fig:slab}
Schematic illustration of the supercell geometries for the two differently
terminated ATiO$_3$ (001) surfaces.}
\end{figure}

We have studied both types of surface termination for all
three materials (PbTiO$_3$, BaTiO$_3$, and SrTiO$_3$)
using a repeated slab geometry.  The slabs are symmetrically
terminated and typically contain seven layers (17 or 18 atoms),
as illustrated in Fig.~\ref{fig:slab}. The vacuum region was chosen
to be two lattice constants thick. The calculations were done using a
(4,4,2) Monkhorst-Pack mesh,\cite{mesh} corresponding to three or
four k-points in the irreducible Brillouin zone for cubic and
tetragonal surfaces respectively. The convergence of the calculations
has been very carefully checked for PbTiO$_3$ by repeating some of
the calculations with asymmetrically terminated eight-layer slabs and
symmetrically terminated nine-layer slabs. Additionally, we have
enlarged the vacuum region to a thickness of three lattice constants,
and we have checked the convergence of the Brillouin zone integration
by going to a (6,6,2) k-point mesh. In all cases, the results for
the structural properties of the surfaces given in the Tables
\ref{tab:relaxI} to \ref{tab:fedist} change by less than 0.2\%.

For all three materials, we first computed the relaxations
for the ``cubic'' surface, i.e., for the case where there is no
symmetry lowering relative to a slab of ideal cubic material.
In this case we preserved $M_x$, $M_y$, and $M_z$ mirror
symmetries relative to the center of the slab, and set the
lattice constants in the $\widehat{x}$ and $\widehat{y}$ directions
equal to those computed theoretically for the corresponding
bulk material (3.89~\AA, 3.95~\AA, and 3.86~\AA\ for PbTiO$_3$,
BaTiO$_3$, and SrTiO$_3$, respectively).  The symmetry-allowed
displacements of the atoms in the $z$ (surface-normal) direction
were then fully relaxed.

Each of the three materials studied displays a different sequence
of structural phase transitions from the cubic paraelectric
phase as the temperature is lowered.\cite{lines}
PbTiO$_3$ undergoes a single transition into a tetragonal ferroelectric
(FE) phase at 763\,K and then remains in this structure down to zero
temperature.
BaTiO$_3$ displays a series of three transitions to tetragonal,
orthorhombic, and rhombohedral FE phases at 403\,K, 278\,K, and 183\,K,
respectively.
SrTiO$_3$ remains cubic down to 105\,K, at which point it undergoes an
antiferrodistortive transition involving rotation of the oxygen
octahedra and doubling of the unit cell.  The material nearly
goes ferroelectric at about $T=30$\,K, but is evidently prevented
from doing so by quantum zero-point fluctuations.\cite{zhong-qsr}

Because we are primarily interested in the room-temperature
structures of these materials and their surfaces, we have
chosen to focus on the tetragonal FE phases of PbTiO$_3$
and BaTiO$_3$ for our surface studies.
We consider only the case of the tetragonal $c$ axis (i.e.,
polarization) lying {\it parallel} to the surface, since polarization
normal to the surface is strongly suppressed by the depolarization
fields that would arise from the accumulated charge at the
surfaces.\cite{zhong-loto}  We take the tetragonal axis to lie
along $\widehat{x}$, and relax the $M_x$ symmetry while retaining the
$M_y$ and $M_z$ symmetries with respect to the center of the
slab.  For PbTiO$_3$, which is tetragonal at $T=0$, this will
indeed be the ground-state structure of the slab.  For
BaTiO$_3$, on the other hand, the $M_y$ symmetry is artificially
imposed so that the theoretical $T=0$ calculation will mimic
the experimental room-temperature surface structure.
In both cases, the slab lattice constants in the
$\widehat{x}$ and $\widehat{y}$
directions were set equal to the corresponding theoretical
equilibrium lattice constants computed for the bulk tetragonal
phase:
$c$=4.04~\AA\ and $a$=3.86~\AA\ for PbTiO$_3$, and
$c$=3.99~\AA\ and $a$=3.94~\AA\ for BaTiO$_3$.

% ------------------
\subsection{Surface Energies}
\label{sec:surfenergy}
% ------------------

A comparison of the relative stability of the AO and TiO$_2$
surface terminations is problematic because the corresponding
surface slabs contain different numbers of AO and TiO$_2$
formula subunits.  We treat this problem by introducing
chemical potentials $\mu_{\rm AO}$ and $\mu_{\rm TiO_2}$
for these subunits, defined in such a way that
$\mu_{\rm AO}=0$ and $\mu_{\rm TiO_2}=0$ correspond to
thermal equilibrium with bulk crystalline AO and TiO$_2$,
respectively.  We have computed the cohesive energies
$E_{\rm AO}$ and $E_{\rm TiO_2}$ of crystalline
PbO, BaO, SrO, and TiO$_2$ using the same first-principles
pseudopotential method in order to provide these reference
values.  The grand potential for a given surface structure
can then be computed as
\begin{equation}
F_{\rm surf} = {1\over 2} [ E_{\rm slab}
       - N_{\rm TiO_2} ( E_{\rm TiO_2} + \mu_{\rm TiO_2} )
       - N_{\rm AO} ( E_{\rm AO} + \mu_{\rm AO} ) ] \;,
\end{equation}
where $N$ is the number of formula subunits contained in the
slab, and the factor of $1/2$ accounts for the fact that each
slab contains two surfaces.  Assuming that the surface
of the ATiO$_3$ is in equilibrium with its own bulk, it
follows that
\begin{equation}
 \mu_{\rm AO} + \mu_{\rm TiO_2} = -E_{\rm f}  \;,
\end{equation}
where $E_{\rm f}$ is the heat of formation of bulk
ATiO$_3$ from bulk AO and bulk TiO$_2$.  The two chemical
potentials are thus not independent, and we choose to treat
$\mu_{\rm TiO_2}$ as the independent variable when presenting
our results.  Accordingly, $\mu_{\rm TiO_{2}}$ is allowed to vary over
the range
\begin{equation}
 -E_{\rm f} \le \mu_{\rm TiO_{2}}\le 0 \;,
\end{equation}
the lower and upper limit corresponding to the precipitation of
particulates of AO and TiO$_2$ on the surface, respectively.

% ------------------------------------------------------
\section{Results and Discussions}
\label{sec:results}      
% ------------------------------------------------------

% -------------------
\subsection{Structural relaxations}
\label{sec:structure}      
% -------------------

We begin by presenting our new results on the structural properties of the
PbTiO$_3$ (001) surfaces. The equilibrium atomic positions for both surface
terminations in the two phases were obtained by starting from the ideal
structures of the surfaces and then relaxing the atomic positions while
preserving the symmetries described in section \ref{sec:geometry}. The results
for the fully relaxed geometries are summarized in Table \ref{tab:relaxI} and
\ref{tab:relaxII}. By symmetry, there are no forces along $\widehat{x}$ and
$\widehat{y}$ for the cubic surface, and no forces along $\widehat{y}$ for the
tetragonal surface.

Tables \ref{tab:relaxI} and \ref{tab:relaxII} show for both surfaces
a substantial inward contraction towards the bulk for the uppermost
surface layers, whereas for the second layers we find an outward relaxation of
the atoms relative to the positions of the atoms on the ideal surface.
Generally, the metal and the oxygen atoms move in the same direction,
but the relaxations of the metal atoms are much larger, leading to a
rumpling of the layers. The single exception is the surface layer of the
tetragonal TiO$_2$--terminated surface, where one of the two oxygen atoms
moves in the opposite
direction to the metal atom. Therefore we can see here a significant asymmetry
between the O atoms with respect to their positions perpendicular to the
surface. This asymmetry between the oxygen atoms in the topmost surface layer
of the tetragonal
TiO$_2$--terminated surface was also found for BaTiO$_3$ but with a
much smaller amplitude. As expected, we find the largest relaxations for the
surface--layer atoms, but the displacement of the Pb atom in the second layer
of the TiO$_2$--terminated surface is of the same magnitude.

In order to compare these results with previous calculations for SrTiO$_3$
and BaTiO$_3$, we have calculated the changes in the interlayer distances
$\Delta d_{ij}$ and the amplitudes of the rumpling $\eta_i$ of the layers in
the surface slabs for all three perovskites. The results for both surface
terminations and the different phases are given in the Tables \ref{tab:surfI}
and \ref{tab:surfII}. We denote the change in the $z$ position of a
\begin{table}
\begin{center}
\begin{minipage}{0.5\textwidth}
\begin{tabular}{lddd}
Atom & $\delta_z(C)$ & $\delta_x(T)$ & $\delta_z(T)$ \\ \hline
Pb(1)            &  $-$4.36  &   $-$3.44  &  $-$2.38  \\
O$_{\rm III}$(1) &  $-$0.46  &  $+$11.85  &  $-$1.17  \\
Ti(2)            &  $+$2.39  &   $+$3.62  &  $+$1.15  \\
O$_{\rm I}$(2)   &  $+$1.21  &   $+$9.27  &  $+$0.81  \\
O$_{\rm II}$(2)  &  $+$1.21  &  $+$11.45  &  $+$0.06  \\
Pb(3)            &  $-$1.37  &   $+$0.00  &  $-$0.81  \\
O$_{\rm III}$(3) &  $-$0.20  &  $+$11.14  &  $-$0.17  \\
Ti(4)            &     0     &   $+$3.86  &     0     \\
O$_{\rm I}$(4)   &     0     &   $+$9.60  &     0     \\
O$_{\rm II}$(4)  &     0     &  $+$10.98  &     0     \\
\end{tabular}
\end{minipage}
\end{center}
\medskip
\caption{\label{tab:relaxI}
Atomic relaxations (relative to ideal atomic positions) of the
PbO--terminated surface in the cubic ($C$\/) and tetragonal ($T$\/) phases.
The relaxations perpendicular ($\delta_z$) and parallel ($\delta_x$) to
the surface are given in
percent of the lattice constants $a$ and $c$, respectively. For reference,
the theoretical $\delta_x$ values in the bulk ferroelectric phase,
relative to the Pb atoms, are
$\delta_x({\rm Ti})=3.45$, $\delta_x({\rm O_I})=9.26$ and
$\delta_x({\rm O_{II}})=\delta_x({\rm O_{III}})=10.44$. Atom labels
refer to Figs.~\ref{fig:struc} and \ref{fig:slab}; results are only
given for the top half of the slab, since the bottom half is equivalent
by $M_z$ mirror symmetry.}
\end{table}
\begin{table}
\begin{center}
\begin{minipage}{0.5\textwidth}
\begin{tabular}{lddd}
Atom & $\delta_z(C)$ & $\delta_x(T)$ & $\delta_z(T)$ \\ \hline
Ti(1)            &  $-$3.40  &   $+$3.62  &  $-$3.47  \\
O$_{\rm I}$(1)   &  $-$0.34  &   $+$9.27  &  $-$1.60  \\
O$_{\rm II}$(1)  &  $-$0.34  &  $+$11.45  &  $+$0.79  \\
Pb(2)            &  $+$4.53  &   $+$0.00  &  $+$4.06  \\
O$_{\rm III}$(2) &  $+$0.43  &  $+$11.14  &  $+$0.17  \\
Ti(3)            &  $-$0.92  &   $+$3.86  &  $-$0.79  \\
O$_{\rm I}$(3)   &  $-$0.27  &   $+$9.60  &  $-$0.03  \\
O$_{\rm II}$(3)  &  $-$0.27  &  $+$10.98  &  $-$0.06  \\
Pb(4)            &    0      &   $-$3.44  &     0     \\
O$_{\rm III}$(4) &    0      &  $+$11.85  &     0     \\
\end{tabular}
\end{minipage}
\end{center}
\medskip
\caption{\label{tab:relaxII}
Atomic relaxations of the TiO$_2$--terminated surface in the cubic ($C$\/) and
tetragonal ($T$\/) phases. Notation is the same as in Table \ref{tab:relaxI}.}
\end{table}
metal atom relative to the ideal unrelaxed structure as $\delta_z({\rm M})$,
and $\delta_z({\rm O})$ is the same for the oxygen atom in the same layer
(defined as $[\delta_z({\rm O_I}) + \delta_z({\rm O_{II}})]/2$ for a TiO$_2$
layer). We then define the change of the interlayer
distance $\Delta d_{ij}$ as the difference between the averaged atomic
displacements $[\delta_z({\rm M})+\delta_z({\rm O})]/2$ of layer $i$ and $j$,
and the rumpling $\eta_i$ is defined as the amplitude of these displacements
$|\delta_z({\rm M}) - \delta_z({\rm O})|$.
From Tables \ref{tab:surfI} and \ref{tab:surfII} we can see that, for
all three perovskites and for both terminations, the surfaces display
a similar oscillating
relaxation pattern with a reduction of the interlayer distance $d_{12}$, an
expansion of $d_{23}$ and again a reduction for $d_{34}$. However, compared to
BaTiO$_3$ and SrTiO$_3$, the amplitudes of the relaxations in PbTiO$_3$ are
significantly increased.

\begin{figure}
$\phantom{dummy}$
\vfill
\epsfxsize=0.5\textwidth
\centerline{\epsffile{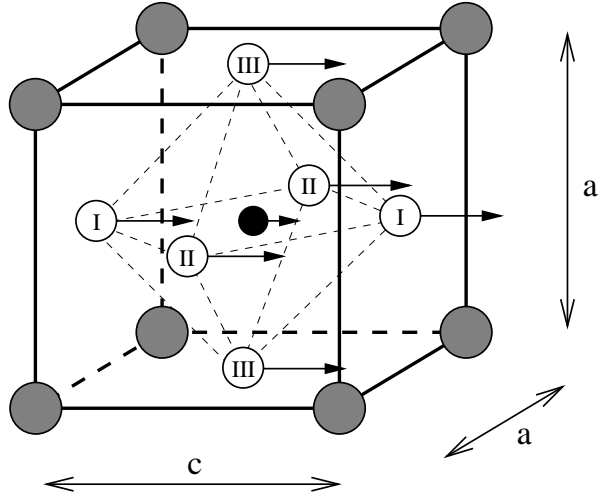}}
\vfill
\caption{\label{fig:struc}
Structure of the cubic perovskite compounds ATiO$_3$. Atoms A, Ti and O
are represented by shaded, solid and open circles,
and O$_{\rm I}$, O$_{\rm II}$
and O$_{\rm III}$ are the oxygen atoms lying along the $\widehat{x}$,
$\widehat{y}$ and $\widehat{z}$
direction from the Ti atom, respectively. Arrows indicate the
displacements of the Ti and O atoms relative to the A atoms in the case of
the tetragonal phase of PbTiO$_3$.}
\end{figure}

\begin{table}
\begin{center}
\begin{minipage}{0.6\textwidth}
\begin{tabular}{l|d|dd|dd}
   &  SrTiO$_3$  &  \multicolumn{2}{c|}{BaTiO$_3$}  &
\multicolumn{2}{c}{PbTiO$_3$} \\
\parbox{28pt}{\hfill} &
\parbox{42pt}{\centerline{cubic}} &
\parbox{36pt}{\centerline{cubic}} &
\parbox{36pt}{\centerline{tetrag}} &
\parbox{36pt}{\centerline{cubic}} &
\parbox{36pt}{\centerline{tetrag}} \\ \hline
$\Delta d_{12}$ &  $-$3.4  &  $-$2.8  &  $-$2.8  &  $-$4.2  &  $-$2.6  \\
$\Delta d_{23}$ &  $+$1.2  &  $+$1.1  &  $+$1.1  &  $+$2.6  &  $+$1.3  \\
$\Delta d_{34}$ &  $-$0.6  &  $-$0.4  &  $-$0.4  &  $-$0.8  &  $-$0.5  \\
$\eta_1$        &     5.8  &     1.4  &     1.5  &     3.9  &     1.2  \\
$\eta_2$        &     1.2  &     0.4  &     0.5  &     1.2  &     0.7  \\
$\eta_3$        &     1.1  &     0.3  &     0.4  &     1.2  &     0.6  \\
\end{tabular}
\end{minipage}
\end{center}
\medskip
\caption{\label{tab:surfI}
Change of the interlayer distance $\Delta d_{ij}$ and layer rumpling
$\eta_i$ (in percent of the lattice constant $a$\/) for the relaxed
AO--terminated surface of the three perovskites in the cubic and tetragonal
phases.}
\end{table}

\begin{table}
\begin{center}
\begin{minipage}{0.6\textwidth}
\begin{tabular}{l|d|dd|dd}
   &  SrTiO$_3$  &  \multicolumn{2}{c|}{BaTiO$_3$}  &
\multicolumn{2}{c}{PbTiO$_3$} \\
\parbox{28pt}{\hfill} &
\parbox{42pt}{\centerline{cubic}} &
\parbox{36pt}{\centerline{cubic}} &
\parbox{36pt}{\centerline{tetrag}} &
\parbox{36pt}{\centerline{cubic}} &
\parbox{36pt}{\centerline{tetrag}} \\ \hline
$\Delta d_{12}$ &  $-$3.5  &  $-$3.1  &  $-$2.9  &  $-$4.4  &  $-$4.1  \\
$\Delta d_{23}$ &  $+$1.6  &  $+$0.9  &  $+$1.2  &  $+$3.1  &  $+$2.5  \\
$\Delta d_{34}$ &  $-$0.6  &  $-$0.6  &  $-$0.4  &  $-$0.6  &  $-$0.4  \\
$\eta_1$ &     1.8  &     2.3  &     2.5  &     3.1  &     3.1  \\
$\eta_2$ &     3.0  &     1.9  &     2.1  &     4.1  &     3.9  \\
$\eta_3$ &     0.2  &     0.4  &     0.4  &     0.7  &     0.8  \\
\end{tabular}
\end{minipage}
\end{center}
\medskip
\caption{\label{tab:surfII}
Change of the interlayer distance $\Delta d_{ij}$ and layer rumpling
$\eta_i$ (in percent of the lattice constant $a$\/) for the relaxed
TiO$_2$--terminated surface of the three perovskites in the cubic and
tetragonal phases.}
\end{table}

The second interesting feature of Tables \ref{tab:surfI} and \ref{tab:surfII}
is that for BaTiO$_3$, there is almost no difference in the relaxations of
the surface layers between the cubic and the tetragonal phase. The same is
true for the TiO$_2$--terminated surface of PbTiO$_3$. For the PbO--terminated
surface, in contrast, the changes in the interlayer distances and the
layer rumplings are strongly reduced in the tetragonal phase. We will come
back to this point at the end of the next subsection.

% ----------------------------
\subsection{Influence of the surface upon ferroelectricity}
\label{sec:fe}
% ----------------------------

We turn now to the question of whether the presence of the surface has a strong
effect upon the near--surface ferroelectricity. To analyze whether the
ferroelectric order is enhanced or suppressed near the surface, we introduce
average ferroelectric distortions $\delta_{\rm FE}$ for each layer of the
surface slabs:

\begin{table}
\begin{tabular}{c|dd|dd|dd|dd}
   &  \multicolumn{4}{c|}{AO--terminated}  &
\multicolumn{4}{c}{TiO$_2$--terminated} \\
   &  \multicolumn{2}{c|}{BaTiO$_3$}  & \multicolumn{2}{c|}{PbTiO$_3$}  &
\multicolumn{2}{c|}{BaTiO$_3$}  & \multicolumn{2}{c}{PbTiO$_3$} \\
layer & $\delta_{\rm FE}$(BaO) & $\delta_{\rm FE}$(TiO$_2$) &
$\delta_{\rm FE}$(PbO) & $\delta_{\rm FE}$(TiO$_2$) &
$\delta_{\rm FE}$(BaO) & $\delta_{\rm FE}$(TiO$_2$) &
$\delta_{\rm FE}$(PbO) & $\delta_{\rm FE}$(TiO$_2$) \\ \hline
 1   &  1.6  &       &  15.3  &       &       &  4.4  &        &  5.7 \\
 2   &       &  1.8  &        &  6.8  &  1.4  &       &   7.0  &      \\
 3   &  1.3  &       &  11.1  &       &       &  3.4  &        &  6.3 \\
 4   &       &  2.6  &        &  6.4  &  1.7  &       &   9.7  &      \\
bulk &  1.5  &  3.2  &  10.4  &  6.4  &  1.5  &  3.2  &  10.4  &  6.4 \\
\end{tabular}
\bigskip
\caption{\label{tab:fedist}
Average layer-by-layer ferroelectric distortions $\delta_{\rm FE}$ of the
relaxed slabs, in percent of the lattice constant $c$\/. Last row shows
the theoretical bulk values for reference.}
\end{table}

\begin{equation}
\begin{array}{rcll}
\delta_{\rm FE} & = &
| \delta_x({\rm A}) -  \delta_x({\rm O_{III}})| &
\mbox{for AO planes and} \\
\delta_{\rm FE} & = &
| \delta_x({\rm Ti}) - [\delta_x({\rm O_{I}}) + \delta_x({\rm O_{II}})]/2|
\qquad &
\mbox{for TiO$_2$ planes.}
\end{array}
\end{equation}
The calculated values of $\delta_{\rm FE}$ for BaTiO$_3$ and PbTiO$_3$ are
given in Table \ref{tab:fedist}; the last row of the table gives the bulk
values for reference.

For the PbO--terminated surface of PbTiO$_3$, one can see a clear
increase in the average ferroelectric distortions $\delta_{\rm FE}$
when going from the bulk values to the surface layer. On the other
hand, for the TiO$_2$--terminated surface, the average distortions are
slightly decreased at the surface.  Surprisingly, this is just the
opposite of what one observes for BaTiO$_3$, where one sees a
reduction of the ferroelectric distortions for the BaO--terminated
surface and a moderate enhancement for the TiO$_2$--terminated
surface.  (Of course, the distortions are also much smaller for
BaTiO$_3$ surfaces, as they are in the bulk, compared to PbTiO$_3$.)
These results tend to confirm that Pb is a much more active
constituent in PbTiO$_3$ than is Ba in BaTiO$_3$, presumably because
of the partially covalent nature of the Pb--O bonds as discussed
in Sec.~\ref{sec:introduction}.

In any case, the present results again confirm that the presence of
the surface does not lead to any drastic suppression of the ferroelectric
order near the surface, supporting the view that extrinsic effects must
be responsible for degradation of ferroelectricity in thin-film
geometries.

\begin{figure}
$\phantom{dummy}$
\vfill
\epsfxsize=0.6\textwidth
\centerline{\epsffile{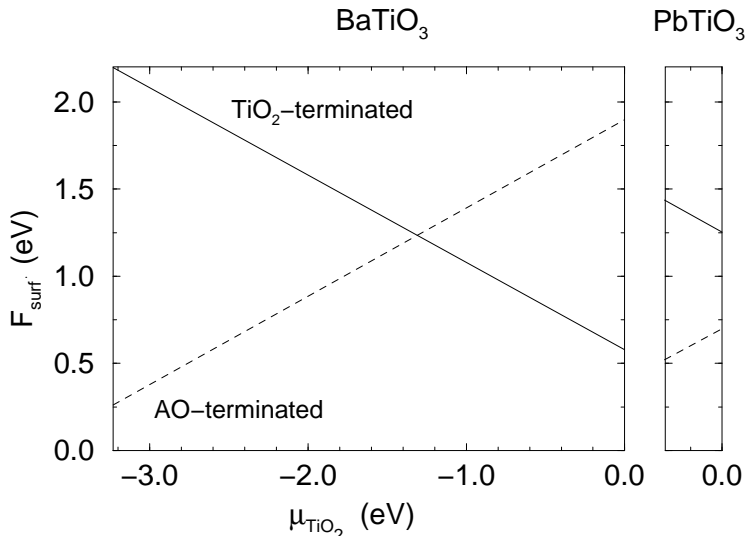}}
\vfill
\caption{\label{fig:fsurf}
Grand thermodynamic potential $F_{\rm surf}$ as a function of the chemical
potential $\mu_{\rm TiO_2}$ for the two types of surfaces of BaTiO$_3$
(left) and PbTiO$_3$ (right), in the tetragonal phase. Dashed and solid
lines correspond to AO--terminated and TiO$_2$--terminated surfaces,
respectively.}
\end{figure}

Finally, we note that there are interesting signs of interplay between
the relaxations parallel and perpendicular to the surface for PbTiO$_3$.
In particular, the relaxations perpendicular to the surface are
substantially reduced (by a factor of $\sim$3) on the PbO-terminated
surface when going from the cubic to the tetragonal case. This can
be rationalized as follows.  Because of the partial
covalency of the Pb--O bonds, there is a tendency to reduce the Pb--O bond
length (this length is 2.75, 2.51, and 2.30 \AA\ in cubic PbTiO$_3$,
tetragonal PbTiO$_3$, and PbO, respectively).
For the cubic surface, by symmetry,
the only possibility to shorten this bond length is by a strong movement of
the Pb atom towards the bulk and a strong movement upwards of the O atoms in
the second layer. This leads to the strong rumpling and the decrease
of $d_{12}$. But in the tetragonal phase there is also the possibility
to enlarge the ferroelectric distortion in order to shorten the Pb--O bond
length. Evidently, the enlargement of the ferroelectric distortion is
preferred to the relaxation perpendicular to the surface.

% ----------------------------
\subsection{Surface energies}
% ----------------------------

In this section we discuss the surface energetics of the three perovskite
compounds. In order to compare the relative stability of the AO-- and 
TiO$_2$--terminated surfaces, we have calculated the grand thermodynamic
potential $F_{\rm surf}$ (as introduced in Sec.~\ref{sec:surfenergy})
for the different surfaces as a function of the chemical potential
$\mu_{\rm TiO_2}$. The results for the tetragonal surfaces of BaTiO$_3$
and PbTiO$_3$ are shown in Fig.~\ref{fig:fsurf}.
The graphs of the grand thermodynamic potentials for the SrTiO$_3$ surfaces
are very similar to those of BaTiO$_3$ and are therefore not shown separately.

Figure \ref{fig:fsurf} shows a very different behavior for the
BaTiO$_3$ and PbTiO$_3$ surfaces. First of all, the formation energy
$E_{\rm f}$ of PbTiO$_3$ (when formed from bulk PbO and TiO$_2$) is
0.36~eV, much lower than the formation energies of SrTiO$_3$ and BaTiO$_3$
which are about 3.2~eV.
This leads to a much smaller range for the chemical potential
$\mu_{\rm TiO_2}$ for which PbTiO$_3$ surfaces can grow in thermodynamic
equilibrium. Second, for BaTiO$_3$ the two different surfaces have a
comparable range of thermodynamic stability, indicating that either
BaO--terminated surfaces or TiO$_2$--terminated surfaces could be formed
depending on whether growth occurs in Ba--rich or Ti--rich conditions.
In contrast,
for PbTiO$_3$ only the PbO--terminated surface can be obtained in
thermodynamic equilibrium.

\begin{table}
\begin{center}
\begin{minipage}{0.6\textwidth}
\begin{tabular}{l|d|dd|dd}
   &  SrTiO$_3$  &  \multicolumn{2}{c|}{BaTiO$_3$}  &
\multicolumn{2}{c}{PbTiO$_3$} \\
\parbox{26pt}{\hfill} &
\parbox{42pt}{\centerline{cubic}} &
\parbox{34pt}{\centerline{cubic}} &
\parbox{34pt}{\centerline{tetrag}} &
\parbox{34pt}{\centerline{cubic}} &
\parbox{34pt}{\centerline{tetrag}} \\ \hline
$E_{\rm f}$     &  3.2   &  3.20  &  3.23  &  0.30  &  0.36  \\
$E_{\rm surf}$  &  1.26  &  1.24  &  1.24  &  0.97  &  0.97  \\
$E_{\rm relax}$ &  0.18  &  0.13  &        &  0.21  &  0.22  \\
\end{tabular}
\end{minipage}
\end{center}
\medskip
\caption{\label{tab:energies}
Formation energy $E_{\rm f}$, average surface energy $E_{\rm surf}$ and
average relaxation energy $E_{\rm relax}$ (in eV/unit cell) for the three
perovskites in the cubic and tetragonal phases.}
\end{table}

To get a quantity describing the surface energetics that is independent of
the chemical potential $\mu_{\rm TiO_2}$ and therefore allows a more direct
comparison of the three compounds, we define the average surface energy
per surface unit cell
\begin{equation}
E_{\rm surf} = \frac{1}{4} \Big( E_{\rm slab}^{\rm AO} +
E_{\rm slab}^{\rm TiO_2} - 7\, E_{\rm bulk} \Big) \;,
\end{equation}
which is equal to the average of the grand thermodynamic potential
$F_{\rm surf}$ for the two kinds of surfaces. Again, the results for
$E_{\rm surf}$ shown in Table \ref{tab:energies} are very similar for
SrTiO$_3$ and BaTiO$_3$, whereas the value for PbTiO$_3$ is significantly
lower.

Finally we have computed the average relaxation energy $E_{\rm relax}$
of the three perovskite compounds. $E_{\rm relax}$ is defined as the
difference between the average surface energy $E_{\rm surf}$ of the ideal
surface without relaxation of the atoms, and the fully relaxed surfaces.
The largest and smallest value for $E_{\rm relax}$
(see Table \ref{tab:energies}) were found for PbTiO$_3$ and
BaTiO$_3$, respectively, which is in agreement with the observation that the
atomic relaxations are largest in PbTiO$_3$ and smallest in BaTiO$_3$.

For all three compounds the average relaxation energy $E_{\rm relax}$ is many
times larger than a typical bulk ferroelectric well depth, which is
approximately 0.03~eV for BaTiO$_3$ and 0.05~eV for PbTiO$_3$. This would
indicate that the surface is capable of acting as a strong perturbation on the
ferroelectric order. As we have shown in Sec.~\ref{sec:fe}, this is not the
case for BaTiO$_3$ and PbTiO$_3$. One reason why the ferroelectric order is
not as strongly affected by the surface as one might have thought has
been pointed out in Ref.~\onlinecite{pad-ba}: the soft phonon eigenmode, which
is responsible for the ferroelectric distortion, is only one of three zone
center modes having the same symmetry. By looking at how strongly the surface
relaxations are related to each of these zone center modes it has turned out
that the distortions induced by the presence of the surface are to a large
extent of non--ferroelectric character.

% ----------------------------
\subsection{Surface band structure}
% ----------------------------

For all three perovskite compounds we have carried out LDA calculations of the
bulk and the surface electronic structure for our various surface slabs. It is
well known that the LDA is quantitatively unreliable regarding excitation
properties such as band gaps. Since we are in the following only looking
at differences between band structures, we think that our conclusions drawn
from the LDA results are nevertheless qualitatively correct.

As has already been shown in Ref.~\onlinecite{king}, the bulk band structures
of SrTiO$_3$ and BaTiO$_3$ are very similar, whereas PbTiO$_3$ shows some
significant differences. In SrTiO$_3$ and BaTiO$_3$ the upper edge of the
valence band is very flat throughout the Brillouin zone. On the other hand,
in PbTiO$_3$ the shallow $6s$ semicore states of the Pb atoms hybridize with
the $2p$ states of the O atoms, leading to a lifting of the upper valence
band states near the X point of the Brillouin zone.

\begin{table}
\begin{center}
\begin{minipage}{0.6\textwidth}
\begin{tabular}{l|d|dd|dd}
   &  SrTiO$_3$  &  \multicolumn{2}{c|}{BaTiO$_3$}  &
\multicolumn{2}{c}{PbTiO$_3$} \\
\parbox{54pt}{\hfill} &
\parbox{40pt}{\centerline{cubic}} &
\parbox{30pt}{\centerline{cubic}} &
\parbox{30pt}{\centerline{tetrag}} &
\parbox{30pt}{\centerline{cubic}} &
\parbox{30pt}{\centerline{tetrag}} \\ \hline
AO-term.      &  1.86  &  1.80  &  2.01  &  1.53  &  2.12  \\
TiO$_2$-term. &  1.13  &  0.84  &  1.18  &  1.61  &  1.79  \\
bulk          &  1.85  &  1.79  &  1.80  &  1.54  &  1.56  \\
\end{tabular}
\end{minipage}
\end{center}
\medskip
\caption{\label{tab:bdgap}
Calculated band gaps (in eV) for the relaxed cubic and tetragonal surface
slabs.}
\end{table}

This fact is responsible for a different behavior of the PbTiO$_3$ surface
band structure compared to SrTiO$_3$ and BaTiO$_3$. If we look at the
calculated band gaps in Table \ref{tab:bdgap}, we see that for
TiO$_2$--terminated surfaces the band gap is significantly reduced for
SrTiO$_3$ and BaTiO$_3$, whereas for PbTiO$_3$ the band gap is almost
unchanged. The reduction of the band gap in SrTiO$_3$ and BaTiO$_3$ is mainly
due to an upward intrusion of the upper valence band states near the M point
into the lower part of the band gap (as pointed out in
Ref.~\onlinecite{pad-ba}, this is caused by the suppression of the
hybridization of certain O $2p$ and Ti $3d$ orbitals in the surface layer).
In PbTiO$_3$ we find the same upward movement of the upper valence band
states near the M point, but these states stay just below the highest valence
states at the X point, and so the band gap is almost unchanged.

On the other hand, for the AO--terminated surfaces we see no reduction
of the band gap for any of the three perovskite compounds.  Even here,
however, there is a subtle difference between PbTiO$_3$ and the other
materials, this time concerning the conduction band edge.  According
to our calculations, the Pb $6p$ states overlap the Ti $3d$ states to some
degree in bulk PbTiO$_3$, and this effect is accentuated at the $\Gamma$
point of the surface Brillouin zone on the Pb--O terminated surface,
where the lowest Pb $6p$ state falls just below the lowest Ti $3d$ state.
We thus suggest that the conduction band minimum may actually have Pb $6p$
character at this surface, although the effect is too small to
affect the band gaps in Table \ref{tab:bdgap} substantially.
This might be an interesting target of investigation
for future spectroscopic experimental studies.

% ------------------------------------------------------
\section{Summary}
\label{sec:summary}
% ------------------------------------------------------

In summary, we have calculated structural and electronic properties of
PbTiO$_3$ (001) surfaces using a first-principles density-functional
approach.  The results are compared and contrasted with corresponding
previous calculations on BaTiO$_3$ and SrTiO$_3$ surfaces.  We observe
qualitatively different behavior of the PbTiO$_3$ surfaces in several
respects.  First, within the narrow range of PbO and TiO$_2$ chemical
potentials permitted by bulk thermodynamics, we find that the
TiO$_2$-terminated surface is never thermodynamically stable.  Thus,
the PbO-terminated surface is expected to be the one observed
experimentally.  Second, the interaction between the ferroelectric
distortion and the presence of the surface is quite different for
PbTiO$_3$, compared to BaTiO$_3$.  In particular, the ferroelectricity
is strongly enhanced at the AO-terminated surface and suppressed
at the TiO$_2$-terminated surface, just the opposite of the
behavior found for BaTiO$_3$.  Moreover, the ferroelectric
distortion at the surface allows for a drastic reduction of the
rumpling of the surface layer on the PbO-terminated surface, an effect
which is not seen on the BaO-terminated of BaTiO$_3$.  Third,
the surface electronic band structure is qualitatively modified in
the case of PbTiO$_3$ by the presence of Pb $6s$ and $6p$ states
in the upper valence and lower conduction regions.

% ------------------------------------------------------
\acknowledgments
% ------------------------------------------------------

This work was supported by the ONR grant N00014-97-1-0048.

% ------------------------------------------------------

%-----------------------------------------------------------------------
\end{document}